\documentclass[aps,prb,twocolumn,groupedaddress,showpacs, nobibnotes]{revtex4-1}
\usepackage[dvips]{graphicx}
\usepackage{vmargin}
\setpapersize{USletter}
\setmarginsrb{0.8in}{.6in}{0.8in}{0.8in}{0in}{0.2in}{0in}{0in}
\begin{document}

\title{Two-dimensional quantum percolation with binary non-zero hopping integrals}
\author{Brianna S. Dillon Thomas}\email{dillonb@purdue.edu}
\author{Hisao Nakanishi}\email{hisao@purdue.edu}
\affiliation{Department of Physics, Purdue University, West Lafayette, IN 47907}
\date{\today}

\begin{abstract}
In a previous work [Dillon and Nakanishi, Eur.Phys.J B 87, 286 (2014)], we numerically calculated the transmission coefficient of 
the two-dimensional quantum percolation problem and mapped out in detail the three regimes of localization, i.e., exponentially 
localized, power-law localized, and delocalized which had been proposed earlier [Islam and Nakanishi, Phys.Rev. E 77, 061109 
(2008)]. We now consider a variation on quantum percolation in which the hopping integral ($w$) associated with bonds that 
connect to at least one diluted site is not zero, but rather a fraction of the hopping integral (V=1) between non-diluted sites. We 
study the latter model by calculating quantities such as the transmission coefficient and the inverse participation ratio and find 
the original quantum percolation results to be stable for $w>0$ over a wide range of energy. In particular, except in the 
immediate neighborhood of the band center (where increasing $w$ to just $0.02*V$appears to eliminate localization effects), 
increasing $w$ only shifts the boundaries between the 3 regimes but does not eliminate them until $w$ reaches 10\%-40\% of 
$V$.
\end{abstract}


\maketitle

\section{Introduction}

Quantum percolation (QP) is one of several models used to study quantum transport in disordered systems. Unlike 
classical percolation, the transport of a particle depends not merely on the underlying connectivity of the system, but 
also on quantum interference effects. Thus, even in a completely connected (ie completely occupied) system, a quantum 
particle's wave function may be of finite extent, resulting in very low or zero transmission, depending on such factors as the 
particle's energy or the system’s boundary conditions. 

Quantum percolation also differs from another common model for disordered systems, the Anderson model, both in model design 
and in transport behavior. While both models can be represented by the same basic Hamiltonian (see Eq.~(\ref{eq1})), the type 
of disorder differs. 

\begin{equation}
H = \sum_{<ij>} \epsilon_{i}|i\rangle \langle i|+\sum_{<ij>} V_{ij}|i\rangle \langle j| + h.c
\label{eq1}
\end{equation}

Whereas in the Anderson model the onsite and/or off-diagonal energies are selected from a finite continuous distribution, in the 
quantum percolation model they are selected from a binary distribution of either zero or infinite energy barriers. In site 
percolation, the on-site energy is randomly chosen to be zero (occupied) or infinite (unoccupied); in bond percolation the off-
diagonal hopping energy is randomly chosen to be either 1 (connected) or zero (disconnected). In both cases, the effect is that 
the disordered site is completely isolated from the rest of the system; a quantum particle is completely unable to move from an 
occupied/connected site to an unoccupied/disconnected site. 

Quantum percolation also differs from the Anderson model in its transport behavior. Previous work by Dillon and Nakanishi 
\cite{dillon14} confirmed numerically that the 2D quantum percolation model exhibits a delocalized state at low disorder and 
mapped a detailed phase diagram showing the three regimes (delocalized, power law localized, and exponentially localized) 
proposed earlier by Islam and Nakanishi \cite{islam08}. Other recent works likewise show a delocalized state 
\cite{schubert08,schubert09, gong09, nazareno02, daboul00}. This is in contrast to the Anderson model, which according to 
one-parameter scaling theory should have only exponentially localized states in the thermodynamic limit in $d=2$ dimensions. 
\cite{abrahams79} While some studies have seen non-localized states or weakly localized states in the Anderson model, these 
appear only at or near the band center \cite{eilmes01}. 

To investigate what differences in the nature of the disorder between the Anderson model and the quantum percolation 
model might lead to their differences in transport behavior, we study a modified quantum percolation model in which the binary 
nature of the QP model’s disorder is maintained, while changing the distribution to a finite one that allows tunneling between 
available and “unavailable” sites. If the infinite-energy aspect of QP disorder is more important, we expect that changing to a 
finite energy will result in losing the QP model's characteristic phases, but not if the binary aspect (which is maintained) is the 
more important characteristic. We start from the approach described in Dillon and Nakanishi \cite{dillon14} using the 
quantum percolation Hamiltonian with off-diagonal disorder and zero onsite energy:

\begin{equation}
H = \sum_{<ij>} V_{ij}|i\rangle \langle j| + h.c
\label{eq2}
\end{equation}

\begin{figure}[t]
{\resizebox{2.7in}{!}{\includegraphics[trim=30 0 330 550]{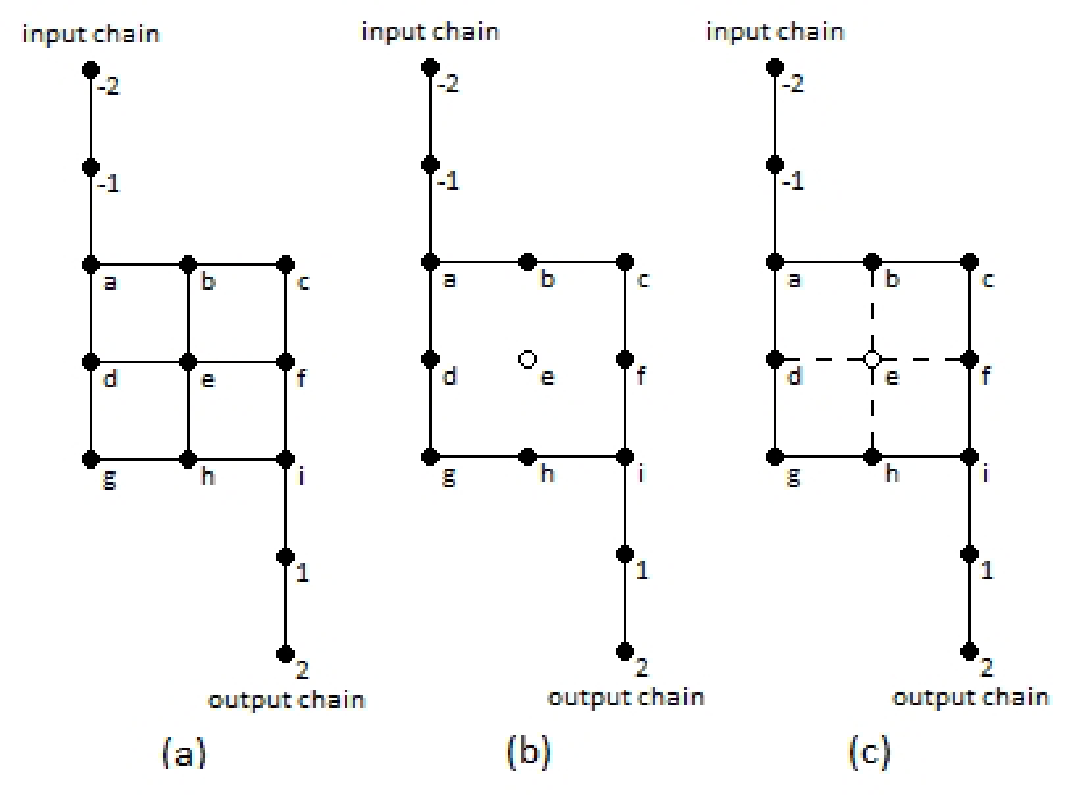}}}\\
\caption{Examples of a small system of $3 \times 3$ square lattice cluster with a point-to-point 
connection. The letters label the lattice points of the cluster part of the Hamiltonian, 
while numbers label those of the leads. Three types of lattices are shown: (a) a fully-connected, ordered lattice, (b) a diluted 
lattice in the original quantum percolation model, and (c) a diluted lattice in the modified quantum percolation model studied in this 
work, with hopping energy $0 \leq q \leq 1$. Lattices (a) and (b) also correspond to the $w=1$ and $w=0$ limits of the modified 
QP model, respectively.}
\label{sq_lat}
\end{figure}

In this study, when randomly diluting the lattice by some fraction $q$, instead of setting $V_{ij}$ = 0 for $i$ and or $j$ 
unoccupied as in the original model, we now set  $V_{ij} = w$,  where $0 \leq w \leq 1$ and $w$ is the same for all diluted sites. 
By doing this, we enable tunneling through and among the diluted sites rather than imposing an infinite energy barrier, while 
still maintaining a binary disorder. As in Ref. \onlinecite{dillon14}, we set up the model on a square lattice of varying sizes to 
which we attach semi-infinite input and output leads at diagonally opposite corners as shown in Fig.~\ref{sq_lat}, and use an 
ansatz proposed by Daboul et al. \cite{daboul00} to calculate the transmission coefficient.  We do this numerically but 
essentially exactly for each realization of the disordered system, and final data are obtained by averaging over anywhere from 
several hundred to a thousand realizations for each dilution $q$, energy $E$, lattice size $L\times L$, and diluted site hopping 
energy $w$. Significantly, we use an identical set of disorder realizations for every diluted site hopping energy $w$ studied for 
each choice of ($q$, $E$, $L$); that is, while the initial selection of disorder realizations for each ($q$, $E$, $L$, $w=0$) is 
random, all subsequent runs for larger values of $w$ at the same ($q$,$E$,$L$) use the same set of disorder realizations.  In 
doing so, we are essentially taking a set of realizations for the original quantum percolation model at a particular ($q$, $E$, $L$) 
and slowly ``turning on" the diluted site hopping energy from $w=0$ to $w=1$ in varying increments. (e.g. in Fig.~\ref{sq_lat}, 
lattices (b) and (c) have the same disorder realization, but in (c) we have ``turned on" the diluted site hopping energy to $w 
\neq 0$). By duplicating the lattice configurations in this manner, we ensure that any differences in transport that 
arise are solely due to changing the hopping energy, not to any differences in the disorder realizations chosen.

The wave function for the entire system of lattice plus input and output leads can be calculated by solving the time-independent 
Schr\"{o}dinger equation:

\begin{equation}
\begin{array}{l}
H\psi = E\psi  \\
\mbox{where}, \psi = \left[ \begin{array}{c} \vec{\psi}_{in} \\ 
                     \vec{\psi}_{cluster} \\ \vec{\psi}_{out} \end{array}
               \right]
\end{array}
\label{eq3}
\end{equation}
and $\vec{\psi}_{in} = \left\{\psi_{-(n+1)}\right\}$ and 
$\vec{\psi}_{out} = \left\{\psi_{+(n+1)}\right\}$,
$n = 0,1,2 \ldots$, are the input and output lead parts of the wave function respectively. The ansatz by Daboul et al. assumes that the input and output parts of the wavefunction are plane waves: 

\begin{equation}
\begin{array}{l}
\psi_{in} {\rightarrow} \psi_{-(n+1)} = e^{-in\kappa} + re^{in\kappa} \\
\psi_{out} {\rightarrow} \psi_{+(n+1)} = te^{in\kappa}
\end{array}
\label{eq4}
\end{equation}

where $r$ is the amplitude of the reflected plane wave and $t$ is the amplitude of the transmitted plane wave. This ansatz  
reduces the infinite-sized problem to a finite one that includes only the main $L \times L$ lattice and the closest points on the 
input and output lead, for the wavevectors ${\kappa}$ that are related to the energy $E$ by: 

\begin{equation}
E = e^{-i\kappa} + e^{i\kappa}
\label{eq5}
\end{equation}

Note that the plane-wave energies in Eq.~\ref{eq5} are restricted to the range $-2 \leq E \leq 2$; by adding the semi-infinite 
one-dimensional input and output leads we have restricted the available energies to the one-dimensional subset of the 
full two dimensional spectrum ($-4 \leq E \leq 4$). However, this range is still large enough for us to observe the localization 
behavior of the wavefunction. 

Having applied the ansatz, the Schr\"{o}dinger equation for an $L \times L$ lattice connected to the semi-infinite input and 
output leads can be reducted to an $(L^{2}+2)\times(L^{2}+2)$ matrix equation of the form:

\begin{equation}
\left[ \begin{array}{ccc}
         -E + e^{i\kappa}  & \vec{c_1}^t  &   0 \\
               \vec{c_1} & \begin{array}[t]{c}
                             {\bf A}
                         \end{array}              &    \vec{c_2} \\
               0     & \vec{c_2}^t  &  -E + e^{i\kappa}
          \end{array} \right]
          \left[ \begin{array}{c}
                    1 + r \\ \vec{\psi}_{clust} \\ t
                    \end{array} \right]
         = \left[ \begin{array}{c}
                  e^{i\kappa} - e^{-i\kappa} \\ \vec{0} \\ 0
                  \end{array} \right]
\label{eq6}
\end{equation}

where {\bf A} is a $L^2 \times L^2$ matrix representing the connectivity of the cluster (with $-E$ as its diagonal elements), 
$\vec{c_i}$ is the $L^2$-component vector representing the coupling of the leads to the corner sites, and $\vec{\psi}_{clust}$ 
and $\vec{0}$ are also $L^2$-component vectors, the former representing the wave function solutions (e.g., $\psi_a$ through 
$\psi_i$ for the cluster in Fig.~\ref{sq_lat}). The cluster connectivity in {\bf A} is represented with $V_{ij}=1$ in positions 
$A_{ij}$ and $A_{ji}$ if sites $i$ and $j$ are connected, otherwise $V_{ij}=w$. For example, the 3x3 lattice shown in 
Fig~\ref{sq_lat}c would have the following matrix equation: 

\begin{widetext}
\begin{equation}
\left( \begin{array}{crrrrrrrrrc}
-E+e^{i\kappa}  &  c  &  0  &  0  &  0  &  0 &  0 &  0 &  0 &  0 &  0 \\
		  c  & -E  &  1  &  0  &  1  &  0 &  0 &  0 &  0 &  0 &  0 \\
		  0  &  1  & -E  &  1  &  0  &  w &  0 &  0 &  0 &  0 &  0 \\
		  0  &  0  &  1  & -E  &  0  &  0 &  1 &  0 &  0 &  0 &  0 \\
		  0  &  1  &  0  &  0  & -E  &  w &  0 &  1 &  0 &  0 &  0 \\
		  0  &  0  &  w  &  0  &  w  & -E &  w &  0 &  w &  0 &  0 \\
		  0  &  0  &  0  &  1  &  0  &  w & -E &  0 &  0 &  1 &  0 \\
		  0  &  0  &  0  &  0  &  1  &  0 &  0 & -E &  1 &  0 &  0 \\
		  0  &  0  &  0  &  0  &  0  &  w &  0 &  1 & -E &  1 &  0 \\
		  0  &  0  &  0  &  0  &  0  &  0 &  1 &  0 &  1 & -E &  c \\
		  0  &  0  &  0  &  0  &  0  &  0 &  0 &  0 &  0 &  c  &  -E+e^{i\kappa} \\
          \end{array} \right)
          \left( \begin{array}{c}
		1 + r \\ \psi_{a} \\ \psi_{b} \\  \psi_{c} \\ \psi_{d} \\ \psi_{e} \\ \psi_{f} \\ \psi_{g} \\ \psi_{h} \\ \psi_{i} \\ t \\
                    \end{array} \right)
         = \left( \begin{array}{c}
                     e^{i\kappa} - e^{-i\kappa} \\ 0 \\ 0 \\ 0 \\ 0 \\ 0 \\ 0 \\ 0 \\ 0 \\ 0 \\ 0
                  \end{array} \right)
\label{eq7}
\end{equation}
\end{widetext}

Eq. ~\ref{eq6} is an exact expression for the 2D system connected to semi-infinite chains with continuous eigenvalues 
between -2 and 2 as specified by the energy restriction above. The solutions of the equation yeild the transmission and 
reflection amplitudes $t$ and $r$, from which we calculate the transmission coefficient $T=|t|^{2}$ and reflection coefficient 
$R=|r|^{2}$.

From the wavefunction solutions of Eq.~\ref{eq6} we also calculate the Inverse Participation Ratio (IPR), which measures 
the fractional size of the particle wavefunction across the lattice and gives a picture of the transport qualities complementary to 
to the picture provided by the transmission coefficient alone. The IPR is defined by: 

\begin{equation}
IPR = \frac{1}{\sum_{i} |\psi_i|^{4} (L^{2})}
\label{eq8}
\end{equation}

where ${\psi_i}$ is the amplitude of the normalized wavefunction for the main-cluster portion of the lattice on site $i$ and $L^2$ 
is the size of the lattice. \cite{comment} It should be noted that our $\vec{\psi_i}$ for given $E$ is a continuum eigenstate of the 
system containing the 1D lead chains, and it is expected to correspond to a mixed state as far as the middle 2E portion is 
concerened. We see that given two lattices of the same size, the one with the smaller IPR has the particle wavefunction residing 
on a smaller number of sites, though the precise geometric distribution cannot be known from the IPR alone. Although the IPR is 
certainly influenced by the lattice's amount of disorder, two different disorder realizations may result in very different IPR, thus 
we calculate not just the average IPR but also the IPR distribution across all realizations. The IPR is often used to assess 
localization by extrapolating it to the thermodynamic limit; if the IPR approaches a constant fraction of the entire lattice, there 
are extended states, whereas if it decays to zero the states are localized. However, various studies have shown that IPR as a 
function of other parameters such as energy can also signal a phase change (see Islam and Nakanishi \cite{nakanishi09}, Johri 
and Bhatt \cite{johri12prl, johri12prb}, and Wang et al. \cite{wang16} for examples in several different systems). 

The remainder of the paper is organized as follows. In Section II, we examine the effects of the diluted site hopping energy on 
the transmission of the modified quantum percolation model, culminating in phase diagrams showing the phase boundary changes 
that result from introducing tunneling. In Section III, we analyze the IPR distributions and averages as a function of both dilution 
and diluted site hopping energy, revealing that the changes induced in the system by introducing tunneling are more complex 
than the transmission coefficient studies alone may suggest. Finally, in Section IV we summarize our results and conclusions.

\section{Transmission and localization} 

The transmission coefficient was calculated over the same 6 energies $E$ (in the range $0.001 \leq E \leq1.6$) and 23 to 27 
lattice sizes $L \times L$ depending on $E$ (where $10 \leq L \leq 780$) as were studied in our previous work, Ref. 
\onlinecite{dillon14}. The dilution range was $2\% \leq q \leq 50\%$  and the diluted-site hopping energy was 
chosen from 23 to 32 values $0 \leq w \leq 1$, both with varying increment sizes. As described in the introduction, the random 
seeding of the disorder realizations was designed such that every $w$ within a given ($q$, $E$,$L$) was applied to an identical 
set of  realizations, while each $q$ (or $E$ or $L$) with the same w was given a different set of realizations, allowing us to 
control for the changes in $w$ while still selecting a random set of lattices over which to average the transmission coefficient. 

\begin{figure*}[htb]
\resizebox{6.7in}{!}{\includegraphics[trim=8 0 0 0]{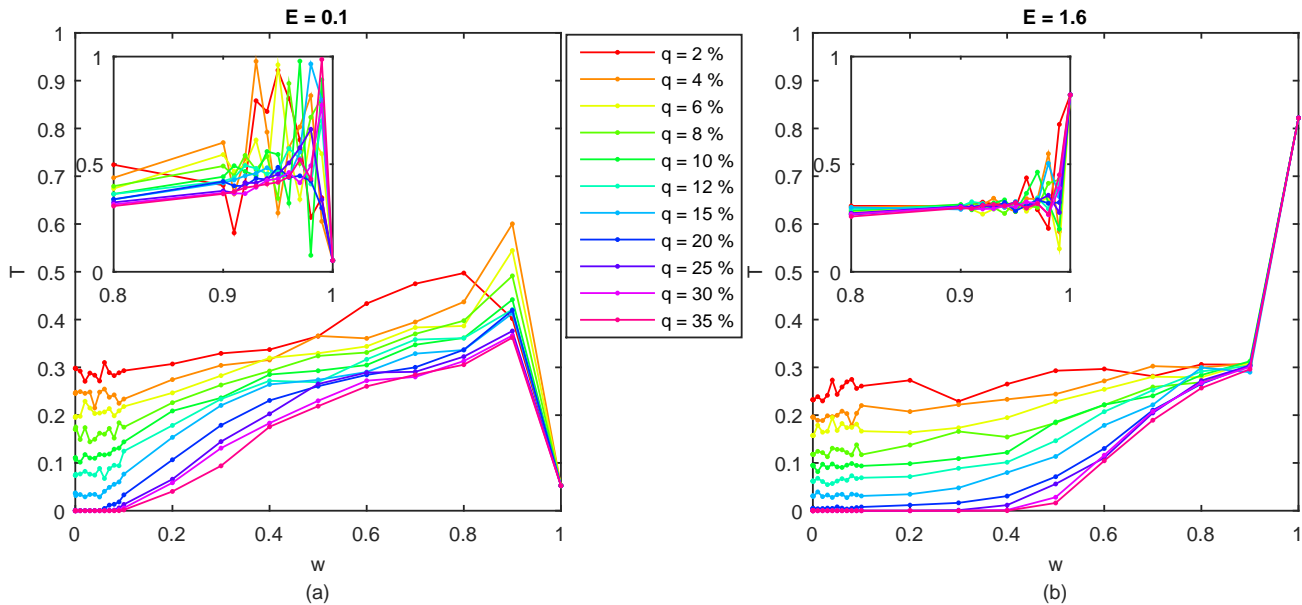}}\\
\caption{(color online) Transmission $T$ vs diluted site hopping energy $w$ on a lattice of size $L=443$ at various dilutions and 
the energies (a)$ E=0.1$ and (b) $E=1.6$, with insets showing a detail of the high-w region with additional points for $w \geq 
0.9$. The lines are merely to guide the eye.}
\label{T_vs_w}
\end{figure*}

Before determining the localization behavior of the modified Hamiltonian in detail, we first examined the transmission $T$ vs the
diluted-site hopping energy $w$ for a few of the larger lattice sizes for each of the energies. An example of two characteristic 
energies away from the band center is shown in Fig.~\ref{T_vs_w}.

\begin{figure*}[htb]
\resizebox{6.7in}{!}{\includegraphics[trim=8 0 0 0]{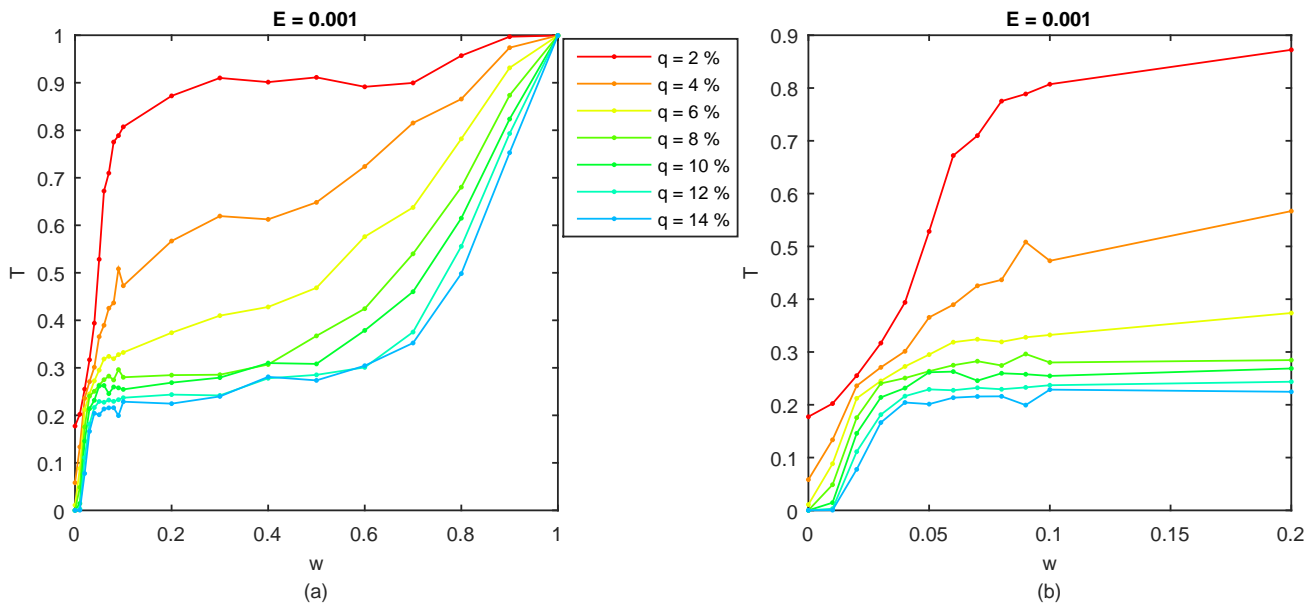}}\\
\caption{(color online) Transmission $T$ vs diluted site hopping energy $w$ on a lattice of size $L=312$ at various dilutions and 
energy $E = 0.001$ for (a) the full range of $w$ studied and (b) zoomed in on $w \leq 0.2$ to more clearly show the initial 
increase in transmisson. The lines are merely to guide the eye.}
\label{T_vs_wcenter}
\end{figure*}

There are two notable features to the transmission curves. Most obviously, there is an abrupt change in the transmission
between $w=0.9$ and $w=1$. Tranmission on an ordered lattice has been shown to depend strongly on the energy, with 
transmission and reflection resonances arising when degenerate eigenstates of the square lattice are split by attaching the 
semi-infinite leads. \cite{cuansing04} Thus, it is not surprising that the $w=1$ limit appears to be a special case. Looking at 
the transmission for smaller increments of $w$ between $w=0.9$ and $w=1$ (see inset in Fig.~\ref{T_vs_w}) shows wide 
fluctuations in the transmission between these two values, thus $w=0.9$ appears to be the lower cut-off for fully-connected-like 
behavior. 

More interesting is the stability of the transmission as $w$ increases. The average fractional cluster size (as measured by simply
counting the fraction of sites for which $|{\psi}|^{2} \neq 0)$ increases from $S \leq q$ for $w=0$ to $S=1$  for $w \geq 
10^{-10}$ (the smallest non-zero hopping energy studied), meaning that at least some tunneling occurs for $w \neq 0$. Despite 
this, the transmission remains stable for up to at least $w = 0.1$ for the smaller energy ($E=0.1$), and as much as $w = 0.3$ 
($E=1.6$), before it increases monotonically to the maximum transmission at $w = 0.9$. 

Near the band center ($E=0.001$, Fig.~\ref{T_vs_wcenter}), the transmission likewise increases monotonically with $w$, but 
does so much more quickly than at higher energies: for all $q$, the transmission increases rapidly for any $w > 0$. Also unique is 
the smooth transition to the fully-connected limit.

\begin{figure*}[htb]
\resizebox{6.7in}{!}{\includegraphics[trim=10 0 0 0]{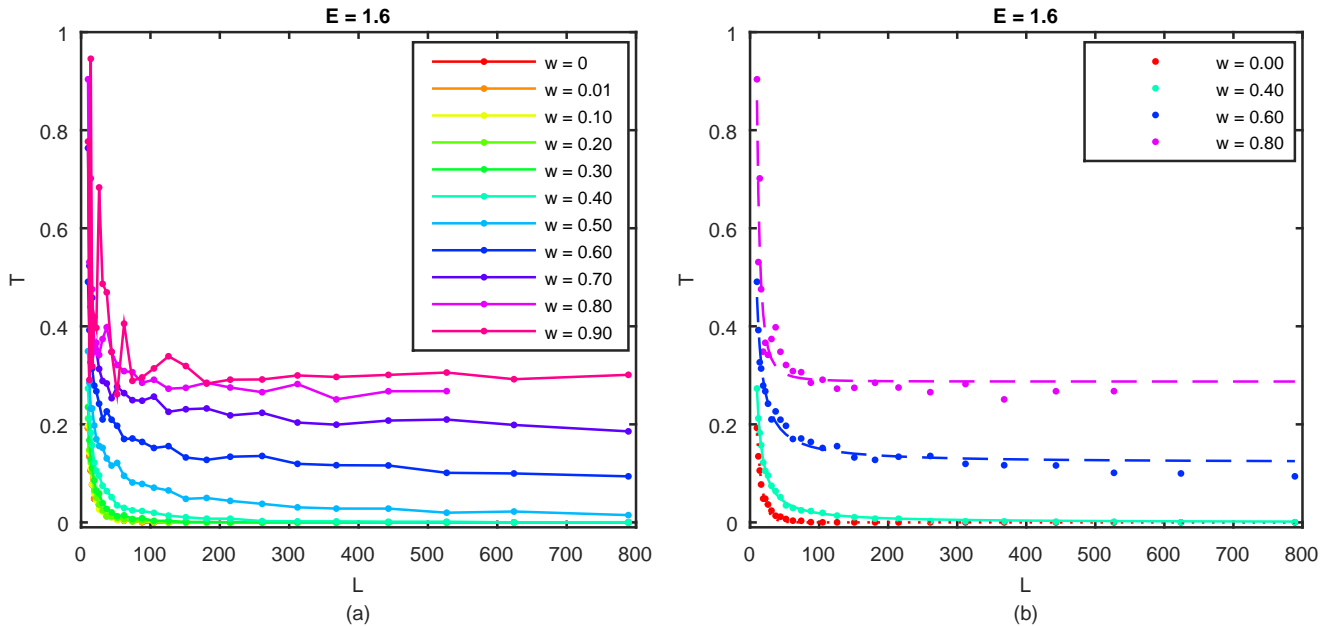}}\\
\caption{(color online) Transmission $T$ vs lattice size $L$ at $E = 1.6$ and q=30\% for selected $w$. In (a) the lines 
connecting the points are to guide the eye; in (b) they are the appropriate fits (dotted: exponential, solid: power law, dashed: 
power law with offset) for the transmission curves, which are a subset of the unfitted curves shown in (a).}
\label{T_vs_L}
\end{figure*}

To construct the complete phase diagram for the modified quantum percolation model, we fit the transmission $T$ vs the lattice
size $L$ for each energy $E$, dilution $q$, and diluted-site hopping energy $w$. As in Ref. \onlinecite{dillon14}, the fit of the 
$T$ vs $L$ curve indicates the state of the system: when an exponential fit ($T=a*exp(-bL)$) is best, it indicates exponential 
localization, a power law fit ($T=aL^{-b}$) indicates a weaker power-law localization, and a fit with an offset (power with offset 
$T = aL^{-b}+c$ orexponential with offset $T=a*exp(-bL)+c$) indicates delocalization since $T=c$ at $L {\rightarrow} 
{\infty}$. For each energy $E$ and dilution $q$, we see the transmission curves progress toward delocalization as the diluted-
site hopping energy
$w$ increases. For $q$ within the exponentially-localized region at $w=0$, the system passes through all three phases: (a) for 
$0 \leq w < w_{p}$  the transmission curve is best fit by an exponential, (b) for $ w_{p} \leq w < w_{d}$ it is best fit by a power 
law, and (c) for $w > w_{d}$ it is best  fit by a power law/exponential plus offset with increasing residual transmission $c$ (see 
Fig ~\ref{T_vs_L} for an example at $E=1.6$). If $q$ is within the power law region at $w=0$, the progression begins at (b) 
with $w_{p}=0$, and for $q$ delocalized at $w=0$ then $w_{d}=0$ and the system only undergoes an increase in residual 
transmission. 

\begin{figure}[!ht]
\hspace{-0.4in}
\resizebox{2.6in}{!}{\includegraphics{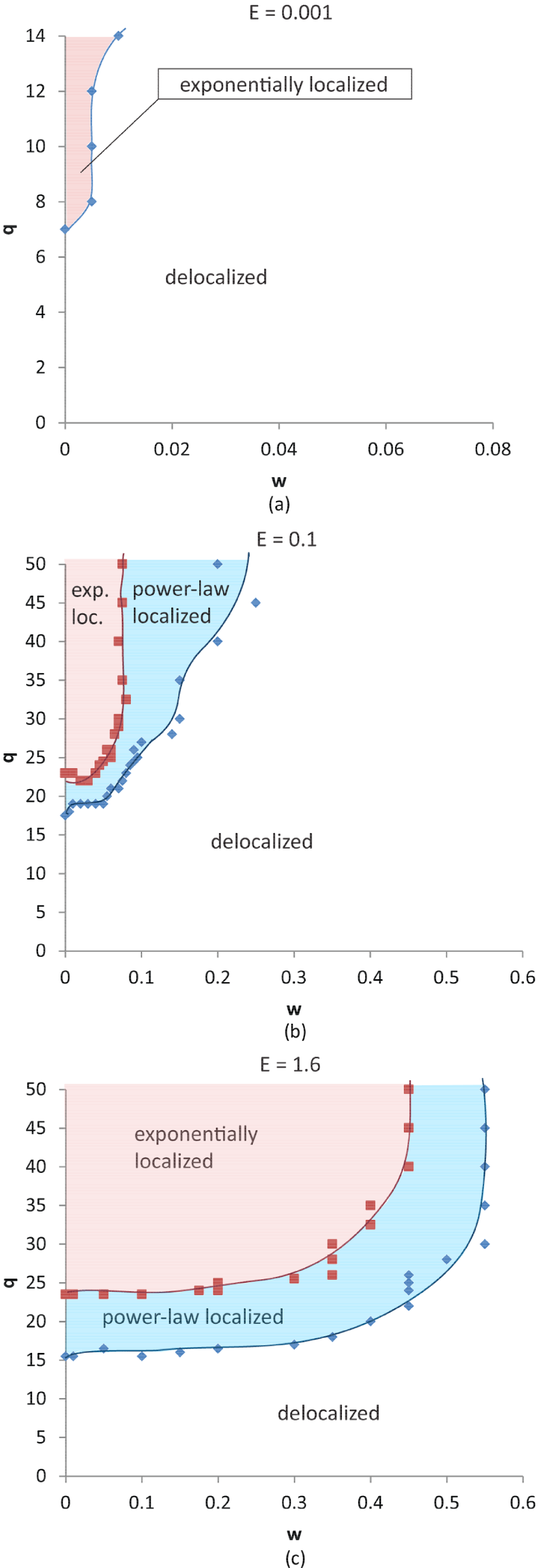}}\\
\caption{(color online) Dilution $q$ vs diluted site hopping energy $w$ phase diagram for the 2D modified quantum percolation 
model at 3 of the 6 energies studied ((a) $E = 0.001$, (b) $E=0.1$, and (c) $E=1.6$). The phase boundaries are to guide the 
eye, not specific fits. The phase diagrams for the energies not shown ($E = 0.4$, $0.7$, and $1.1$) are qualitatively similar to 
those of $E=0.1$ and $E=1.6$, showing the three regions characteristic of quantum percolation with a progressive increase in 
the size of the phase regions from $E=0.1$ to $E=1.6$. $E=0.001$ is a special case in which there is no (or vanishingly small) 
power-law region and the system rapidly becomes delocalized as $w$ increases.}
\vspace{-0.4in}
\label{phase}
\end{figure}

The $q$ vs $w$ phase diagrams for three of the six energies studied are shown in Fig.~\ref{phase}. At $w=0$, the phase 
boundaries match the phase diagram found in Ref \onlinecite{dillon14} within error bars. Near the band center, the exponentially 
localized region seems to vanish for very low $w$. It is possible that for higher dilutions $q$, the exponentially localized region 
persists to larger $w$, however, we did not study these since calculations at $E=0.001$ are extremely computationally 
expensive compared to other energies due to the small diagonal terms making the sparse matrix closer to singular. For all other 
energies, the phase boundaries between the three regions has no quantitative change up to some value $w_{quant}$, with 
$w_{quant}$ as low as $0.05$ for $E=0.1$ and up to $w_{quant}=0.35$ for $E=1.6$. For $w>w_{quant}$, the three regions 
initially persist with the phase boundaries shifted to higher $q$, but as $w$ increases still further the exponentially localized and 
then the power law localized regions disappear, leaving all states delocalized at all dilutions. The transitions from three phases to 
two to only delocalized states each occur at larger $w$ as $E$ increases; the phase diagrams for the three energies not shown 
show a progressive increase in the size of the phase regions from $E=0.1$ to $E=1.6$. For $w=0.6$ the system is delocalized 
at all energies as well as at all dilutions.

The stability of the phase boundaries as $w$ is increased to at least $5\%$ of the available-site hopping energy, combined with 
the presence of the three phases characteristic of quantum percolation for $w$ up to at least $10\%$ and up to as much as 
$40\%$ of the maximum $V = 1$ (depending on energy), lead us to conclude that the binary disorder of the quantum 
percolation model is more significant than the disorder being infinite. Had the latter been more important, we would have 
expected to see the localized phases vanish much more quickly (if not immediately) upon increasing $w$ from 0 (whereby the 
infinite energy barriers associated with diluted sites become finite ones). 

\section{Inverse Participation Ratio}

\begin{figure}[ht]
\resizebox{2.6in}{!}{\includegraphics{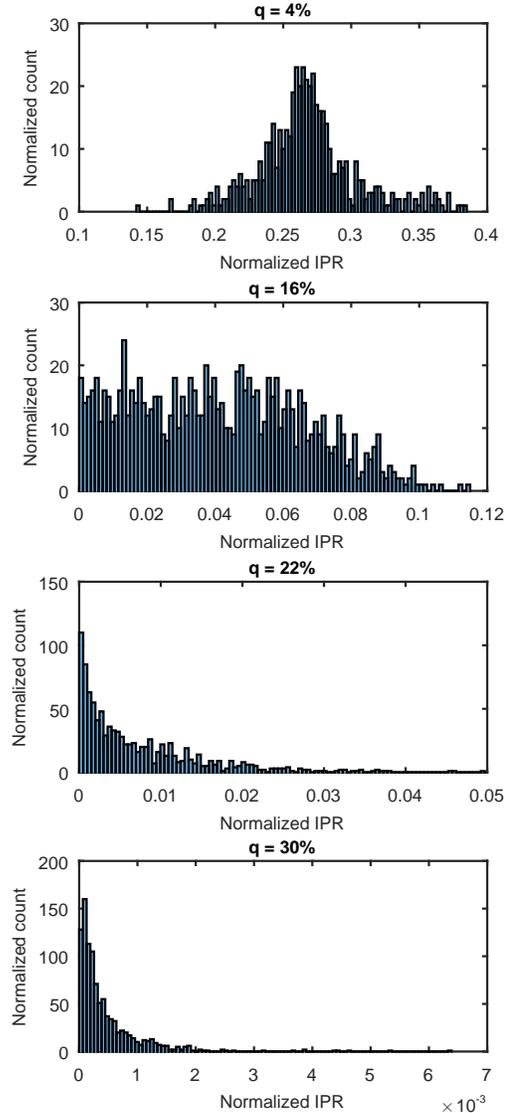}}\\
\caption{(color online) Sample of normalized inverse participation ratio (IPR) histograms at $E = 1.6$ and $L=443$ and $w=0$ 
for four different dilutions within (a) the delocalized region, (b) and (c) the power-localized region, and (c) the exponentially 
localized region, showing the distributions characteristic of each of those phases. (For $E=1.6$ phase boundaries are at 
$q=15.5 \pm 0.5\%$ and $q=24 \pm 1.5\%$.)}
\label{histograms}
\end{figure}

To help us understand why the transmission (and phase) is initially unaffected by particle's nonzero tunneling probability, we look 
at the Inverse Participation Ratio (IPR), first at the $w=0$ limit of the original, unmodified quantum percolation model. For 
$w=0$, the maximum IPR is equal to the percentage of available sites $1-q$; this occurs if the wavefunction is uniformly 
distributed over all available sites. In practice, the IPR will be smaller, due to interference effects and the random application of 
disorder resulting in clusters of theoretically available sites that are disconnected from the main conducting cluster. We examined 
both the IPR distribution across all realizations and the average IPR on the largest available lattice size in common to all dilutions 
at a given energy. At $w=0$, we find that while the average IPR decreases smoothly as $q$ increases, the IPR histogram for 
the disorder realizations exhibits distinct characteristics depending on the transmission state: for delocalized $q$, the IPR is 
peaked and looks roughly Gaussian (Fig.~\ref{histograms}a); for $q$ around and just above the power-law phase boundary, the 
IPR distribution is more box-like with a tail on the right (Fig.~\ref{histograms}b); for higher $q$ within the power law region the 
distribution has a low peak near 0 with a tail (Fig.~\ref{histograms}c); and for exponentially localized $q$, the IPR is strongly 
peaked near 0 with a long tail to the right (Fig.~\ref{histograms}d). Thus the IPR histogram serves as a detailed check 
independent of $L$ on the localization state of the system. 

\begin{figure*}[htb]
\resizebox{6.7in}{!}{\includegraphics{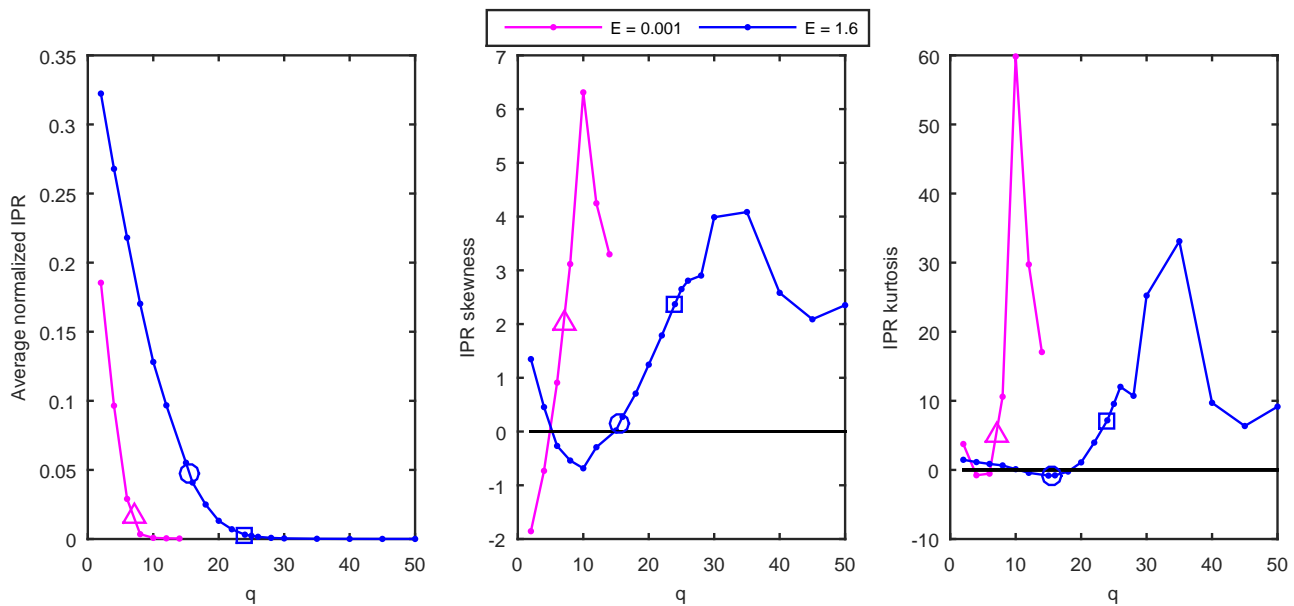}}\\
\caption{(color online) IPR average, skewness, and kurtosis vs dilution $q$ for $E=1.6$ and $E = 0.001$ at $w=0$, the original 
QP model. For $E=1.6$, the delocalized to power-law localized phase boundary and power-law to exponentially localized phase 
boundary are marked on the curve by circles and squares, respectively; for $E=0.001$ the phase boundary between delocalized 
and exponentially localized is denoted by a triangle. }
\label{IPR_vs_Q}
\end{figure*}

Visually checking all histograms for all $E$, $q$, and $w$ is obviously cumbersome, so we instead characterize the distribution by 
its skewness and kurtosis, which measure the symmetry and shape of the distribution, respectively. The skewness of a 
distribution with $n$ elements $x_{i}$ is defined by 

\begin{equation}
Sk = \frac{\frac{1}{n}\sum_{i}(x_{i} - \mu)^{3}}{\sigma^{3}}
\label{eq9}
\end{equation}

where $\mu$ is the mean of $x$ and $\sigma$ is its standard deviation. Skewness = 0 for a symmetric distribution, while positive 
or negative skewness indicates a tail on the right or left  side respectively, and $|Sk|>1$ is generally taken to indicate a 
substantially asymmetrical distribution. The kurtosis of a distribution is defined by 

\begin{equation}
K = \frac{\frac{1}{n}\sum_{i}(x_{i} - \mu)^{4}}{\sigma^{4}}-3\\
\label{eq10}
\end{equation}

where $\mu$ and $\sigma$ are again the mean and standard deviation. Kurtosis = 0 for a normal distribution, negative kurtosis 
indicates a flat, more uniform distribution, and positive kurtosis indicate a strongly peaked distribution. Combining these two 
characteristics with our observations of the histograms in each of the three qualitative phases, we can say that within the 
delocalized phase $|Sk|<1$ and $K \approx 0$ (usually roughly $0 \leq K <1$ because the IPR is slightly more peaked than 
gaussian in delocalized phase), within the power law region $Sk > 0$ with $K < 0$ at the delocalized-to-power-law boundary 
moving to $K>0$ for power-law-to-exponential boundary, and within the exponentially localized phase $Sk>1$ and $K \gg 1$. 
Some examples of the IPR average, skewness, and kurtosis are plotted vs dilution in Fig.~\ref{IPR_vs_Q} at $E=1.6$ and 
$w=0$, with the phase boundaries (as determined from the transmission fits) marked. We see that the phase boundaries 
correspond well with the observed changes in the distribution measures.

\begin{figure*}[htb]
\resizebox{6.7in}{!}{\includegraphics{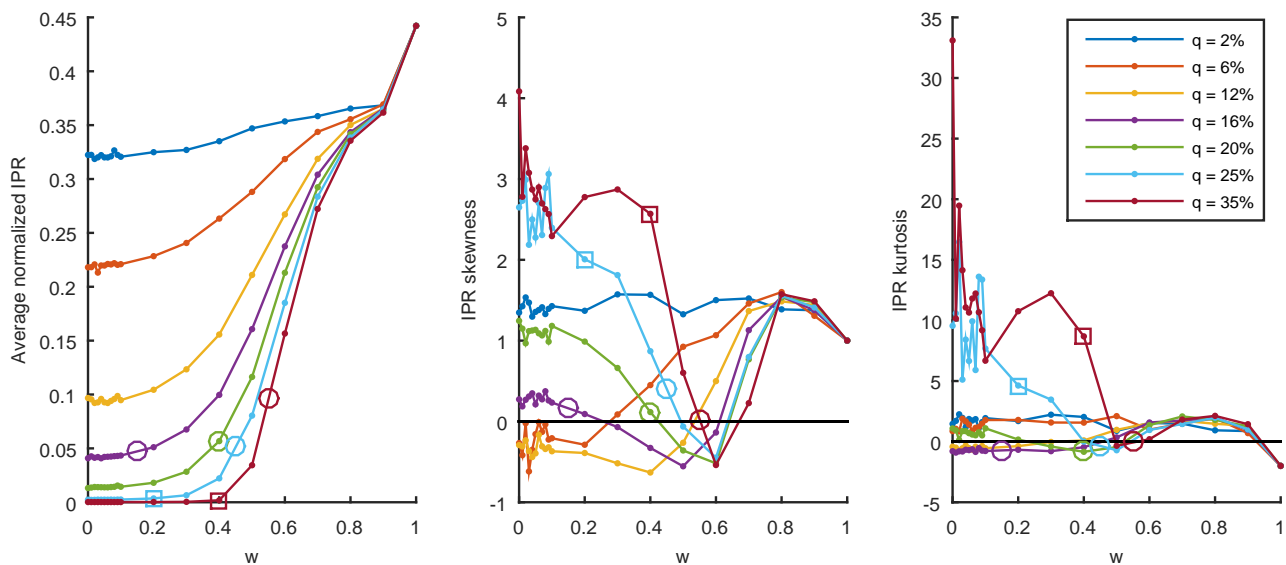}}\\
\caption{(color online) IPR average, skewness, and kurtosis vs diluted site hopping energy $w$ for $E=1.6$ at selected 
dilutions. The exponential to power law localization phase boundary and power law to delocalized phase boundary are denoted 
on each curve by squares and circles, respectively. If a curve has no markers, it is in the delocalized phase for all $w$; if it only 
has one marker, it begins in the power-law localized phase at $w=0$ and shifts into the delocalized phase.}
\label{IPRvsW_E1.6}
\end{figure*}

We then examine the IPR average, skewness, and kurtosis vs $w$ for different $q$, which reveals a more complex picture with 
several interesting features. First, for $E \geq 0.4$ (Fig.~\ref{IPRvsW_E1.6}), we see that for all dilutions, the average IPR, 
skeweness and kurtosis are all roughly constant for for $w \leq 0.1$  except for high $q$ which show fluctuations in 
the kurtosis. The kurtosis is still within the range indicative of a very sharp peak in the IPR, which for high $q$ (localized 
states) is near IPR = 0, therefore we do not believe the fluctuations to be indicative of any significant change in the particle's 
state. Thus for $w \leq 0.1$, the average wavefunction seems essentially locked in place. While we 
know that the particle wavefunction does spread across the entire lattice for $w > 0$ (recall that cluster size = lattice size for 
$w>0$), apparently very little of the wavefunction reaches the newly accessible sites, most likely due to interference effects 
caused by there still bing a strong probability of reflection. Secondly, while the three distinct regions described in the previous 
paragraph are visible in the IPR average, skeweness, and kurtosis combined, the transitions between phases are smooth ones; 
there are no abrupt changes in the wavefunction behavior. As $w$ increases toward the expoenential-to power law phase 
boundary $w_{p}$ (denoted by square markers in Fig ~\ref{IPRvsW_E1.6}) the wavefunction remains exponentially localized 
(average near 0, large skewness, and very large kurtosis indicates a strong peak near IPR=0) but on average begins to spread 
slightly more evenly across the entire lattice including the diluted sites (slight increase in average and decrease in kurtosis means 
fewer realizations with IPR $\approx$ 0). Within the power law region the wavefunction behavior changes the most: in 
increasing $w$ from $w_{p}$ to the power-to-delocalized phase boundary $w_{d}$ (denoted by circle markers in Fig 
~\ref{IPRvsW_E1.6}), the wavefunction continues to spread more evenly across the entire lattice (average IPR increases 
rapidly) while the system shifts smoothly from being dominated by realizations with the wavefunction concentrated on a small 
number of sites (large skewness and kurtosis with smaller IPR)  to realizations with a more uniform mixture of participation ratios 
($sk \approx 0$ with $k < 0$). (The changes within the power law region are true regardless of whether the system began in the 
exponentially localized phase or the power-law localized phase for $w=0$.) Lastly, within the delocalized region ($w > w_{d}$), 
the wavefunction continues to spread more evenly across the entire lattice (average IPR increases), and it becomes more likely 
that different disorder realizations sustain the wavefunction over the same number of sites (skewness and kurtosis indicate a 
shift from a low peak to a tight peak). 

\begin{figure*}[htb]
\resizebox{6.7in}{!}{\includegraphics[trim=0 0 0 -10]{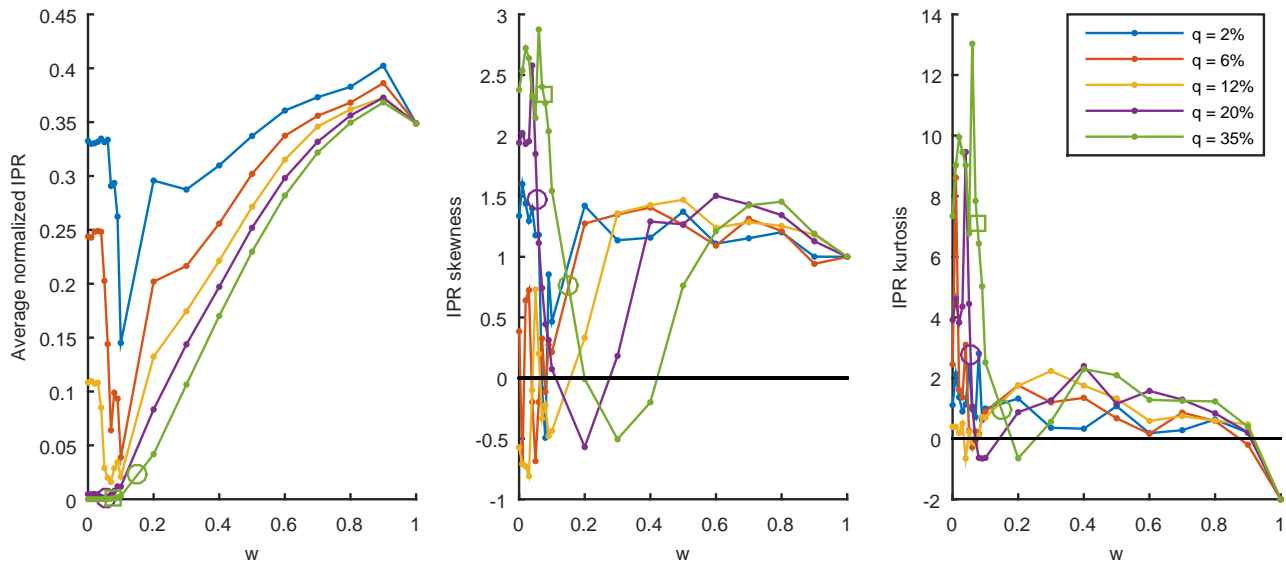}}\\
\caption{(color online) IPR avg, skewness, and kurtosis vs diluted site hopping energy $w$ for $E=0.1$ at selected dilutions. The 
phase boundaries are marked on each curve as in Fig. ~\ref{IPRvsW_E1.6}. We see that there is an anomalous dip in the 
average IPR of initially delocalized $q$ for small $w$.}
\label{IPRvsW_E0.1}
\end{figure*}

For $E=0.1$ and $q \geq 18\%$, the IPR distribution (as described by the average, skewness, and kurtosis) is much the same
as for $E \geq 0.4$, the primary exception being that the distribution remains stable from $w=0$ only up to $w=0.05$, not to 
$w=0.1$ (Fig.~\ref{IPRvsW_E0.1}). However, for $q<18\%$, i.e. dilutions for which the state is delocalized for all $w$, there is 
a curious drop in the average IPR around $w=0.1$ that is not accompanied by a similar drop in the transmission, which is roughly 
constant over this range (compare Fig.~\ref{T_vs_w}). The trend of the skewness and kurtosis of the IPR distribution at 
$E=0.1$ are not very different from the other energies (the phase shifts occur at lower $w$ as given by the phase diagram 
Fig.~\ref{phase}b, but the shape is the same), so it seems the change is only in the average IPR, not in the shape of the 
distribution. The dip around $w=0.1$ occurs for all lattice sizes, making it less likely that it is a finite size effect. While we are not 
certain what could be causing such (apparently) anomalous behavior, our best estimate is again that interference effects are at 
play: it is conceivable that increasing the hopping energy temporarily increases the probability of destructive interference on 
sites that had always been available, thereby constraining the wavefunction to a narrower path while sustaining the same 
transmission, before increasing the hopping energy to a degree that such interference effects are overcome. 

\begin{figure*}[htb]
\resizebox{6.7in}{!}{\includegraphics[trim=0 0 0 -10]{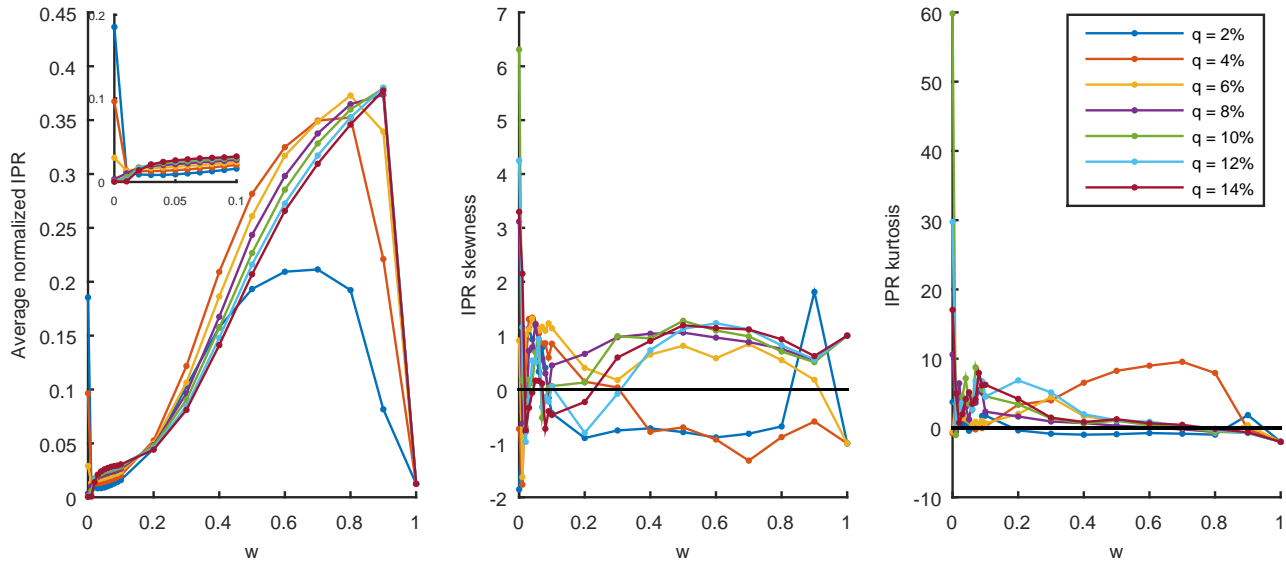}}\\
\caption{(color online) IPR average, skewness, and kurtosis vs diluted site hopping energy $w$ for $E=0.001$ at selected 
dilutions. The phase transitions are not marked on these curves; for $q \geq 8\%$ there is a transition from exponentially 
localized to delocalized at $w=0.005$. For the initially delocalized $q$ there is a drop in the average IPR above $w=0$ (see 
inset), similar to that seen at $E=0.1$. For all dilutions, below $w = 0.2$ the relationship between IPR and $q$ is inverse of 
what we would expect, with the lower dilutions having a lower, not higher, IPR.}
\label{IPRvsW_E0.001}
\end{figure*}

For $E=0.001$, the energy nearest to the band center, we see even more interesting and unexpected behavior. That the IPR is 
small for low $w$ is not entirely unexpected, since the transmission is likewise small, however, unlike at other energies for which 
the IPR and transmission increase at a similar rate, here the IPR increases more slowly than the transmission does. Apparently, 
the wavefunction is constrained to a very narrow selection of sites even at low $q$, so increasing the hopping energy of the 
diluted  sites along that path reduces the destructive interference they cause, thus significantly increasing the transmission even 
as the transmitting cluster remains roughly the same size. This is consistent with our knowledge that for quantum percolation, the 
eigenstates of $E=0$ are are dominated by many state with small spatial extent, leading to lower transmission and localization at 
smaller $q$ in the original model. Our work examines a continuous spectrum in which the particle wavefunction is a mixed state, 
but it is reasonable to believe the mixed state to be similarly dominated by small spatial clusters. More unusual is that for $q=2\% 
- 6\%$ (for which the system is delocalized for all $w$), we again see an unexpected drop in IPR just above $w=0$, where 
there is no such drop in the transmission (see inset on Fig ~\ref{IPRvsW_E0.001}). In this case, due to the increments in $w$ 
studied, it is highly probable that there is actually a singularity in the average IPR at $w=0$; regardless, the behavior is still 
puzzling. Furthermore, for $w \approx < 0.15$, the relationship between IPR and $q$ is inverse of what we would expect (and 
inverse of the relationship between transmission and $q$), that is, we see the smaller dilutions having the lower IPR, meaning 
the wavefunction is more tightly constrained in the lattice at lower dilutions than at higher ones. Again, we are unsure why this is 
the case.

\section{Summary and Discussion}

We have studied a modified version of the quantum percolation model in which the diluted site hopping integrals are allowed to 
be non-zero, thus introducing the possibility of tunneling through and among the previously inaccessible diluted sites while 
maintaining a binary disorder. Our work is based on a system where one-dimensional leads are connected to the diagonal corners 
of a randomly diluted square lattice, for which we calculate the transmission coefficient, inverse participation ratio, and related 
quantities numericall and analyze the results by methods such as finite-size scaling. We determined a full three-parameter phase 
diagram showing the effects of changing the diluted site hopping energy along with the dilution and particle energy. From these 
phase diagrams, we see that the quantum percolation model is a surprisingly robust one, with the three phases characteristic of 
the original model persisting to at least $w=0.05$ for $E \geq 0.1$, and even higher for larger $E$. By examining the average 
participation ratio, we see that in fact, for $w < 0.05$ to $0.1$, the modified model shows results that are nearly identical to the 
original even as far as the individual realizations, with the wavefunction being predominately constrained to the original 
(undiluted) despite a small fraction of the wavefunction tunnelling to and through the now-accessible diluted sites. For these 
very small $w$, then, there are still strong interference effects that continue to work in conjunction with the underlying disorder 
to cause localization. At higher values of $w$, the wavefunction is able to spread more evenly across the entire lattice, but it is 
not until $w \simeq 0.6$, a surprisingly large hopping integral, that the wavefunction is delocalized for all energies $E$. Thus, we 
see that for lower $w$, the modified QP model is dominated by the amount of disorder $q$; though interference effects are 
weakened as the hopping integral increases, it is not enough to affect localization character. Since the quantitative and then 
qualitative characteristics of the original QP model are maintained for such a wide range of diluted site hopping integral, we 
conclude that the binary nature of the disorder is the defining characteristic of the QP model, not the existence of infinite-energy 
barrier. Finally, at sufficiently high $w$, the phase behavior is dominated more by the diluted site hopping integral than the 
amount of disorder $q$ present, evidenced by the vertical phase boundaries for the localized states. 

The energy nearest the band center is the exception to the rule. In this case, we find that increasing the diluted site hopping 
integral quickly moves the system into the delocalized phase for all $q$.  Furthermore, the increased transmission corresponding 
to the phase change is not accompanied by a commensurate change in the IPR, which remains very small and does not increase 
dramatically until $w > 0.2$. It may be possible to interpret these two results combined as reflecting the fact that the 
wavefunctions are constrained to a large number of small spatial clusters at the band center for a wide range of $q$, and thus 
an increase in $w$ has a large effect on suddenly creating a connected path through the lattice for the quantum particle, while it 
has a much smaller effect on creating a large cluster on which the wavefunctions can reside. Perhaps more puzzling is the fact 
that there is a singularity in the IPR at $w=0$ for low values of $q$ that correspond to the delocalized region, and that for $w < 
0.2$ it is the lower dilutions that have a smaller IPR. These peculiarities occur within the range of $w$ for which strong 
interference effects are evident for other energies. Additional study will be needed to interpret this behavior more clearly. 
Regardless, the overal result is consistent with the band center being a special case for the QP model.

We additionally observed an anomalous decrease in the IPR for $E=0.1$ at lower dilutions for which the system is always in the 
delocalized phase. The anomaly only occurs in the average IPR; the shape of the distribution of IPR realizations appears 
unaffected, as is the transmission. We are unsure of what causes this anomaly, or whether it is of significance since it does not 
affect the overall phase of the system.

In conclusion, we have seen that the quantum percolation model is a robust and complex one. That the model remains 
quantitatively unchanged for a range of $w \neq 0$ broadens its applicability to materials and systems in which it is 
unrealistic for impurities to be modeled as completely isolated from the rest of the material. Additionally, we have shown that the 
same phase boundaries found by calculating the transmission can be determined by examining the dependence of the IPR 
distribution  on dilution or hopping integral, giving us a method of assessing the localization properties of the QP model without 
requiring explicit use of finite size scaling and numerical fitting procedures. We expect this to  be useful when studying 
anisotropic lattices or other lattice configurations in which the appropriate extrapolation method for the thermodynamic limit is 
not necessarily clear.

\section*{Ackowledgements}

We thank Purdue University and its Department of Physics for their generous support including 
computing resources.


\begin{thebibliography}{99}
\bibitem{dillon14} B. S. Dillon and H. Nakanishi, Eur. Phys.J B {\bf 87}, 286 (2014)
\bibitem{islam08} M.F. Islam and H. Nakanishi, Phys. Rev. E, {\bf 77}, 061109 (2008)
\bibitem{schubert08} G. Schubert and H. Fehske, Phys. Rev. B {\bf 77}, 245130 (2008)
\bibitem{schubert09} G. Schubert and H. Fehske, in {\it Quantum and 
                 Semi-classical Percolation and Breakdown in Disordered Solids}, 
                 edited by A. K. Sen, K. K. Bardhan, and B. K. Chakrabarti, Springer, pages 163-189 (2009)
\bibitem{gong09} L. Gong and P. Tong, Phys. Rev. B {\bf 80}, 174205 (2009)
\bibitem{nazareno02} H. N. Nazareno, P. E. de Brito and E.S. Rodrigues, Phys. Rev. B {\bf 66}, 012205 (2002)
\bibitem{daboul00} D. Daboul, I. Chang, A. Aharony, Eur. Phys. J B {\bf 16}, 303 (2000)
\bibitem{eilmes01} A. Eilmes, R. A. R\"{o}mer, and M. Schreiber, Physica B, {\bf 296}, 46 (2001)
\bibitem{abrahams79} E. Abrahams, P.W. Anderson, D.C. Licciardello and T.V. Ramakrishnan,  Phys. Rev. Lett., {\bf 42}, 673 (1979)
\bibitem{comment} For our model, we have chosen to normalize the IPR by the lattice size rather than connected cluster size (as is sometimes done) since the lattice size is the fixed parameter and doing so allows better comparison between different sizes when extrapolating to the thermodynamic limit. 
\bibitem{nakanishi09} H. Nakanishi and Md F. Islam, in {\it Quantum and Semi-classical Percolation and Breakdown in Disordered Solids}, edited by A. K. Sen, K. K. Bardhan, and B. K. Chakrabarti, Springer, pages 109-133 (2009)
\bibitem{johri12prl} S. Johri and R. N. Bhatt, Phys. Rev. Lett. {\bf 109}, 076402 (2012)
\bibitem{johri12prb} S. Johri and R. N. Bhatt, Phys. Rev. B {\bf 86}, 125140 (2012)
\bibitem{wang16} Y. Wang, H. Hu, S. Chen, Eur. Phys. J B, {\bf 89}, 77 (2016)
\bibitem{cuansing04} E. Cuansing and Hisao Nakanishi, Phys. Rev. E, {\bf 70}, 066142 (2004)
\end{thebibliography}
\end{document}